\documentclass[journal=jacsat,manuscript=article]{achemso}

\usepackage[version=3]{mhchem} 
\usepackage{textgreek}

\author{Mari Napari}
\affiliation[CU Mat]
{Department of Materials Science and Metallurgy, University of Cambridge, Cambridge CB3 0FS, UK}
\altaffiliation{Present address: Zepler Institute for Photonics and Nanoelectronics, University of Southampton, Southampton SO17 1BJ, UK}
\email{m.p.napari@soton.ac.uk}
\author{Tahmida N. Huq}
\affiliation[CU Mat]
{Department of Materials Science and Metallurgy, University of Cambridge, Cambridge CB3 0FS, UK}
\author{David J. Meeth}
\affiliation[CU EENg]
{Electrical Engineering Division, Department of Engineering, University of Cambridge, Cambridge CB3 0FA, UK}
\author{Mikko J. Heikkil{\"a}}
\affiliation[HelsinkiALD]
{Department of Chemistry, University of Helsinki, FI-00014 Helsinki, FIN}
\author{Kham M. Niang}
\affiliation[CU EENg]
{Electrical Engineering Division, Department of Engineering, University of Cambridge, Cambridge CB3 0FA, UK}
\author{Han Wang}
\affiliation[Purdue]
{Materials Engineering, Purdue University, IN 47907, US}
\author{Tomi Iivonen}
\affiliation[HelsinkiALD]
{Department of Chemistry, University of Helsinki, FI-00014 Helsinki, FIN}
\altaffiliation{Present address: Nanoform Finland Oyj, FI-00790  Helsinki, FIN}
\author{Haiyan Wang}
\affiliation[Purdue]
{Materials Engineering, Purdue University, IN 47907, US}
\author{Markku Leskel{\"a}}
\affiliation[HelsinkiALD]
{Department of Chemistry, University of Helsinki, FI-00014 Helsinki, FIN}
\author{Mikko Ritala}
\affiliation[HelsinkiALD]
{Department of Chemistry, University of Helsinki, FI-00014 Helsinki, FIN}
\author{Andrew J. Flewitt}
\affiliation[CU EENg]
{Electrical Engineering Division, Department of Engineering, University of Cambridge, Cambridge CB3 0FA, UK}
\author{Robert L. Z. Hoye}
\affiliation[CU Mat]
{Department of Materials Science and Metallurgy, University of Cambridge, Cambridge CB3 0FS, UK}
\altaffiliation{Present address: Department of Materials, Imperial College London, London SW7 2AZ, UK}
\author{Judith L. MacManus-Driscoll}
\affiliation[CU Mat]
{Department of Materials Science and Metallurgy, University of Cambridge, Cambridge CB3 0FS, UK}

\title[Role of ALD \ce{Al2O3} surface passivation on the performance of p-type \ce{Cu2O} TFTs]
  {Role of ALD \ce{Al2O3} surface passivation on the performance of p-type \ce{Cu2O} thin film transistors}

\abbreviations{IR,NMR,UV}

\begin{document}







\begin{abstract}

High-performance p-type oxide thin film transistors (TFTs) have great potential for many semiconductor applications. However, these devices typically suffer from low hole mobility and high off-state currents. We fabricated p-type TFTs with a phase-pure polycrystalline \ce{Cu2O} semiconductor channel grown by atomic layer deposition (ALD). The TFT switching characteristics were improved by applying a thin ALD \ce{Al2O3} passivation layer on the \ce{Cu2O} channel, followed by vacuum annealing at 300~$^{\circ}$C. Detailed characterisation by TEM-EDX and XPS shows that the surface of \ce{Cu2O} is reduced following \ce{Al2O3} deposition and indicates the formation of 1--2~nm thick \ce{CuAlO2} interfacial layer. This, together with field-effect passivation caused by the high negative fixed charge of the ALD \ce{Al2O3}, leads to an improvement in the TFT performance by reducing the density of deep trap states as well as by reducing the accumulation of electrons in the semiconducting layer in the device off-state.

\end{abstract}

\section{Introduction}

Metal-oxide thin film transistors (TFTs) have attracted increasing interest especially in display technologies owing to their optical transparency and high mobility, low processing temperatures and material costs, and mechanical flexibility \cite{Hosono2012}. This has led to the development of high-performance n-type semiconducting oxide materials, such as amorphous indium-gallium-zinc-oxide (IGZO) with electron mobility of several tens of cm$^2$V$^{-1}$s$^{-1}$ \cite{Sheng2019}. However, the full utilisation of oxides in p-n junction based electronics and complementary metal oxide semiconductor (CMOS) integrated circuits, is still hindered by the lack of high performance p-type oxides. The reason for the challenges in achieving feasible hole conductivity are the differences in the electronic structures of the n- and p-type oxides \cite{Wang2016}. The transport path of holes in p-type oxides, valence band maximum (VBM) consists typically of localised anisotropic oxygen 2p orbitals, which results in large hole effective mass and low mobility. In addition, the concentration of holes in oxides is often limited by the high formation energy of the cation vacancies, as well as the annihilation of holes due to the low formation energy of the oxygen vacancies \cite{Wang2016}. In case of cuprous oxide \ce{Cu2O}, however, the valence band is formed by the hybridisation of the O 2p and Cu 3d orbitals, resulting in a less localized VBM and pathway for hole transportation for holes formed via copper vacancies (V\textsubscript{Cu}) as acceptor states \cite{Nolan2006}. Such special configuration and high hole mobility have made \ce{Cu2O} an extensively studied p-type oxide for TFTs \cite{AlJahwari2015}, and due to to its other advantageous properties, such as material abundance and solar absorbance, it has also been investigated as a potential candidate for multiple device applications ranging from photovoltaics to sensors \cite{Zhang2019}. 

\ce{Cu2O} layers for TFTs are traditionally fabricated by physical vapour deposition (PVD) methods, such as pulsed lased deposition (PLD) \cite{Matsuzaki2008,Zou2010,Ran2015} and sputtering \cite{Fortunato2010,Jeong2013,Han2016}. Solution-based processing methods have also been used, such as spin coating \cite{Kim2013,Jang2016}, electrodeposition \cite{Musselman2012} and inkjet printing \cite{Baby2015}.  For scalable device applications it is crucial to be able to deposit films with uniform and controllable thickness and composition over large areas, preferably at low or moderate temperatures. Atomic layer deposition (ALD) has been proven invaluable for the fabrication of modern microelectronics, where it is used to produce ultra-thin high-quality dielectric films for devices including metal-oxide semiconductor field effect transistors (MOSFET) and dynamic random access memories (DRAM). ALD has the potential to be extended in production of active device layers. It has already shown to be capable of depositing n-type semiconducting films, such as IGZO\cite{Cho2019}, with properties compatible with what has been achieved by PVD techniques\cite{Sheng2019,Cho2019}. The successful application of ALD grown n-type semiconducting oxides in TFTs has been demonstrated both on rigid and flexible substrates \cite{Sheng2018,Sheng2018_2}. Development of ALD processes for p-type materials (NiO, CuO\textsubscript{x}, SnO) has been mostly of interest for photovoltaics, especially in perovskite and tandem solar cells, where they can be used as electron-blocking and hole transport layers \cite{Zardetto2017}. However, some examples of other electronics applications, such as  p-type TFTs with ALD grown semiconductor channels have been published\cite{Sheng2018,Maeng2016,Kim2017}. For example, high performing TFTs with ALD grown CuO\textsubscript{x} films (consisting of both \ce{Cu2O} and \ce{CuO} phases) have been reported by Maeng et al. \cite{Maeng2016}. Their devices showed an unusually high field effect mobility of \textmu\textsubscript{FE} =  5.6~cm$^2$V$^{-1}$s$^{-1}$, which is higher than the \textmu\textsubscript{FE} of any reported CuO\textsubscript{x} device in the literature. Unfortunately, to our best knowledge, these results have not yet been consistently reproduced, nor are there other reports of the use of ALD \ce{Cu2O} in TFTs. However, ALD was used to demonstrate high-performance p-type TFTs with SnO channel \cite{Kim2017}. There it was observed that applying an \ce{Al2O3} channel passivation significantly improves the TFT performance via the reduction of trap states at the interface.

Here, we investigate the influence of an ALD \ce{Al2O3} passivation layer on the performance of p-type TFTs with an ALD grown \ce{Cu2O} channel. We show that passivation and subsequent vacuum annealing improve the transistor performance metrics. In addition to device measurements, the \ce{Al2O3}/\ce{Cu2O} interface was characterised in detail by using x-ray photoelectron spectroscopy (XPS) and transmission electron microscopy (TEM) to obtain more information of the surface reactions taking place during the deposition of the \ce{Al2O3} on the \ce{Cu2O} and how the passivation layer and the oxide interface affect device performance.

\section{Results and discussion}

\subsection{\ce{Cu2O} film characterization}

\begin{figure}[ht!]
    \centering
    \includegraphics[width=8.5cm]{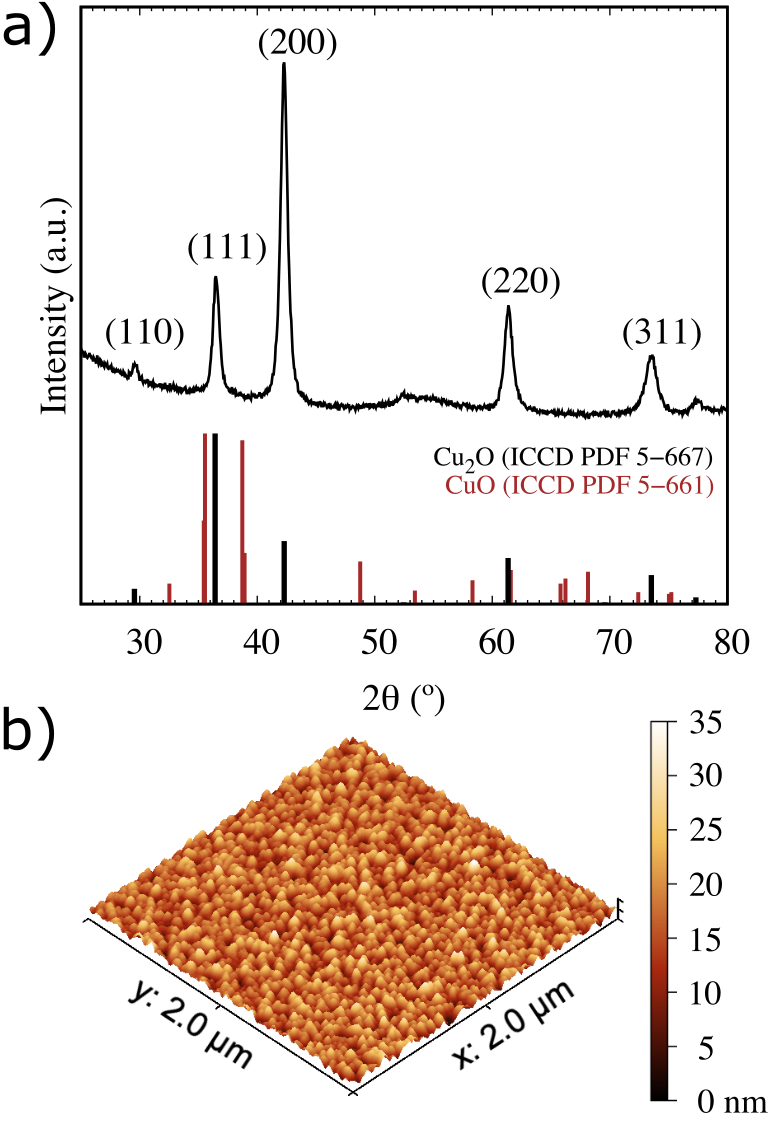}
    \caption{a) XRD pattern of the as-deposited, phase-pure polycrystalline \ce{Cu2O} film and the ICCD cards for both \ce{Cu2O} and CuO, used for indexing. b) 3D AFM image of 2~\textmu m $\times$ 2~\textmu m area of the corresponding film. The film roughness (RMS) is 4.5~nm.}
    \label{Fig1}
\end{figure}

The X-ray diffraction pattern of a 40~nm thick \ce{Cu2O} film is presented in Fig. \ref{Fig1}(a). The GIXRD revealed that the films were polycrystalline \ce{Cu2O}, with the most intense reflections associated to the (200), (111), and (220) planes of the cubic \ce{Cu2O}. No trace of \ce{CuO} or Cu were detected, indicating that the films were phase-pure \ce{Cu2O}, with crystallite size of ca. 30~nm. The crystalline structure of the films was visible also by AFM (see example Fig. \ref{Fig1}(b)) showing the films to have distinct grains in the morphology with a high surface roughness of ca. 4.5~nm (RMS). Despite the high film roughness, we can assume the films to be continuous, based on the detailed growth analysis of corresponding ALD \ce{Cu2O} films reported by Iivonen et al. in Ref. \cite{Iivonen2019}. Hall effect measurements confirmed the p-type conductivity of the films, with a resistivity of \textrho = 300~\textOmega cm, hole density of N = 10$^{16}$~cm$^{-3}$ and Hall mobility \textmu\textsubscript{H} = 0.6~cm$^2$V$^{-1}$s$^{-1}$. The Hall hole mobility is somewhat lower than what has been reported earlier for \ce{Cu2O} films. However, as-deposited films processed at lower temperatures generally pose a lower hole mobility, in the order of few cm$^2$V$^{-1}$s$^{-1}$\cite{AlJahwari2015,Fortunato2010,Han2016,Munos-Rojas2012,Chen2014} as a maximum, than films deposited and/or treated at high temperatures, in which the mobility can reach tens of cm$^2$V$^{-1}$s$^{-1}$\cite{Wang2016,Matsuzaki2008} but the variation between different reports is vast. In our case the low hole mobility may be due to low film thickness, which, combined with small grain size and high surface roughness, limits the conduction. Han et al. have investigated the role of the \ce{Cu2O} film morphology on the charge carrier characteristics, and they concluded that nanocrystalline structure of thin \ce{Cu2O} films can suggest the presence of potential energy barriers at grain boundaries, leading to effects such as grain boundary scattering, which hinders the hole transport in the thin films \cite{Han2017}. This is further enhanced by the formation of a conductive \ce{CuO} layer onto the grain surfaces \cite{Deuermeier2018}. 

\subsection{TFT performance}

\begin{figure}
    \centering
    \includegraphics[width=8.5cm]{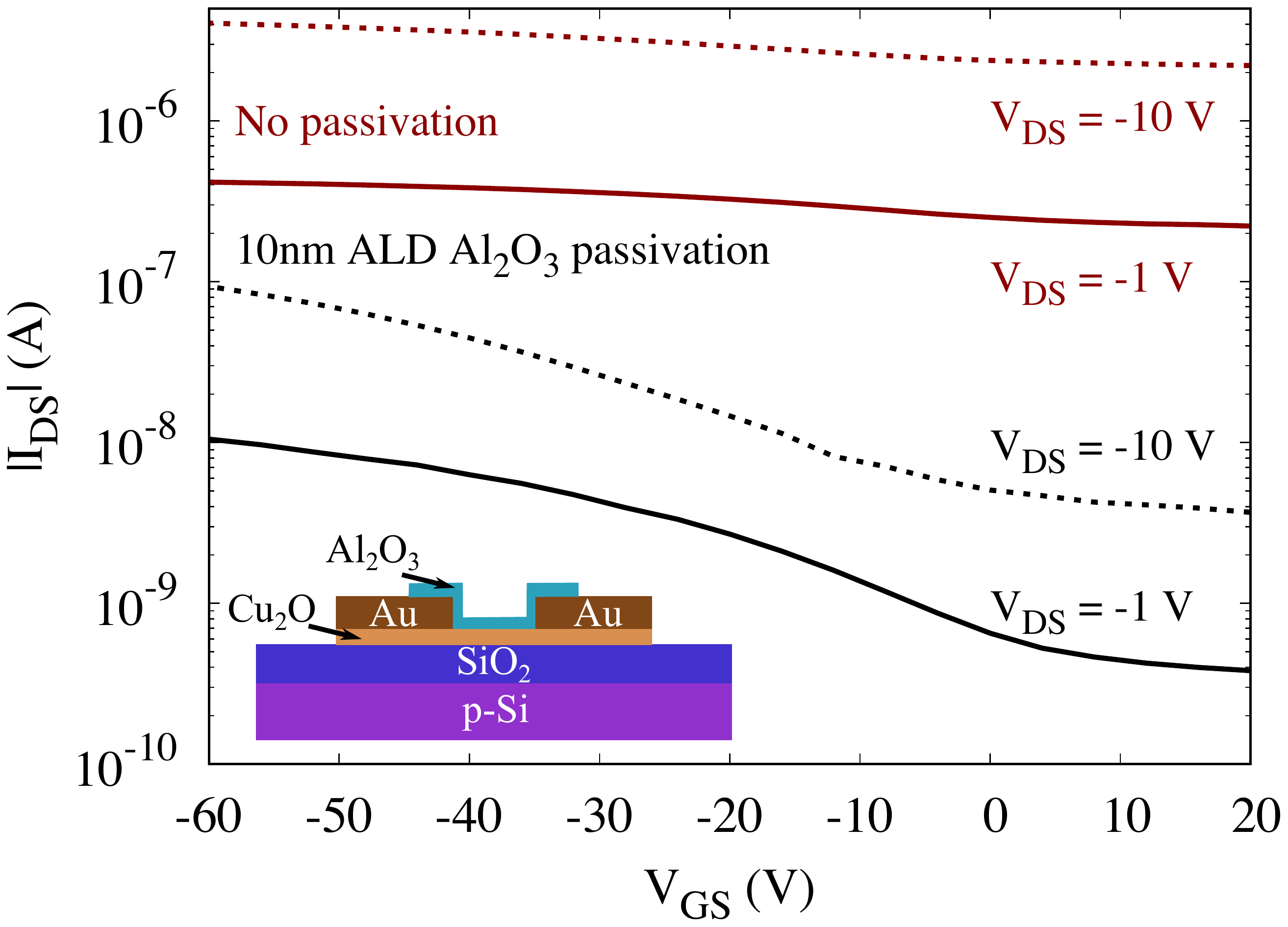}
    \caption{Gate transfer characteristics of a TFT device with 40~nm ALD \ce{Cu2O} p-channel, with and without \ce{Al2O3} passivation, shown as black and dark-red curves, respectively. In both cases the device is measured with drain voltages (V\textsubscript{DS}) of -10.0~ V (dashed lines) and -1.0~V (solid lines). Inset shows the schematic of the TFT device with \ce{Al2O3} passivation layer.}
    \label{Fig2}
\end{figure}

The \ce{Cu2O} films were tested as p-channels in simple bottom-gate thin film transistor devices with Au source and drain electrodes and p-Si substrate acting as a common gate (see inset in Fig. \ref{Fig2}). The switching characteristics of the as-deposited films without the \ce{Al2O3} passivation layer were negligible as shown in Fig. \ref{Fig2}. With a 10~nm \ce{Al2O3} layer deposited on the \ce{Cu2O} channel the off-state drain current ($\vert$I\textsubscript{DS}$\vert$) at positive gate voltage V\textsubscript{GS} decreased by three orders of magnitude and switching with I\textsubscript{on}/I\textsubscript{off}$\approx$ 30 was measured. It has been shown that the gap state density in oxide semiconductor TFTs can be affected by the ambient moisture and oxygen adsorption on the top channel surface, which can be suppressed by the passivation layer \cite{Chen2010}.  However, for this effect the type or fabrication method of the passivation layer seems not to be critical, as improvements in the performance of n- and p-type TFTs have been reported with different ALD and solution-processed oxide films as well as with organic passivation layers \cite{Kim2017,Hu2018,Hong2017,Tak2020}.   

\begin{figure}[ht!]
    \centering
    \includegraphics[width=8.5cm]{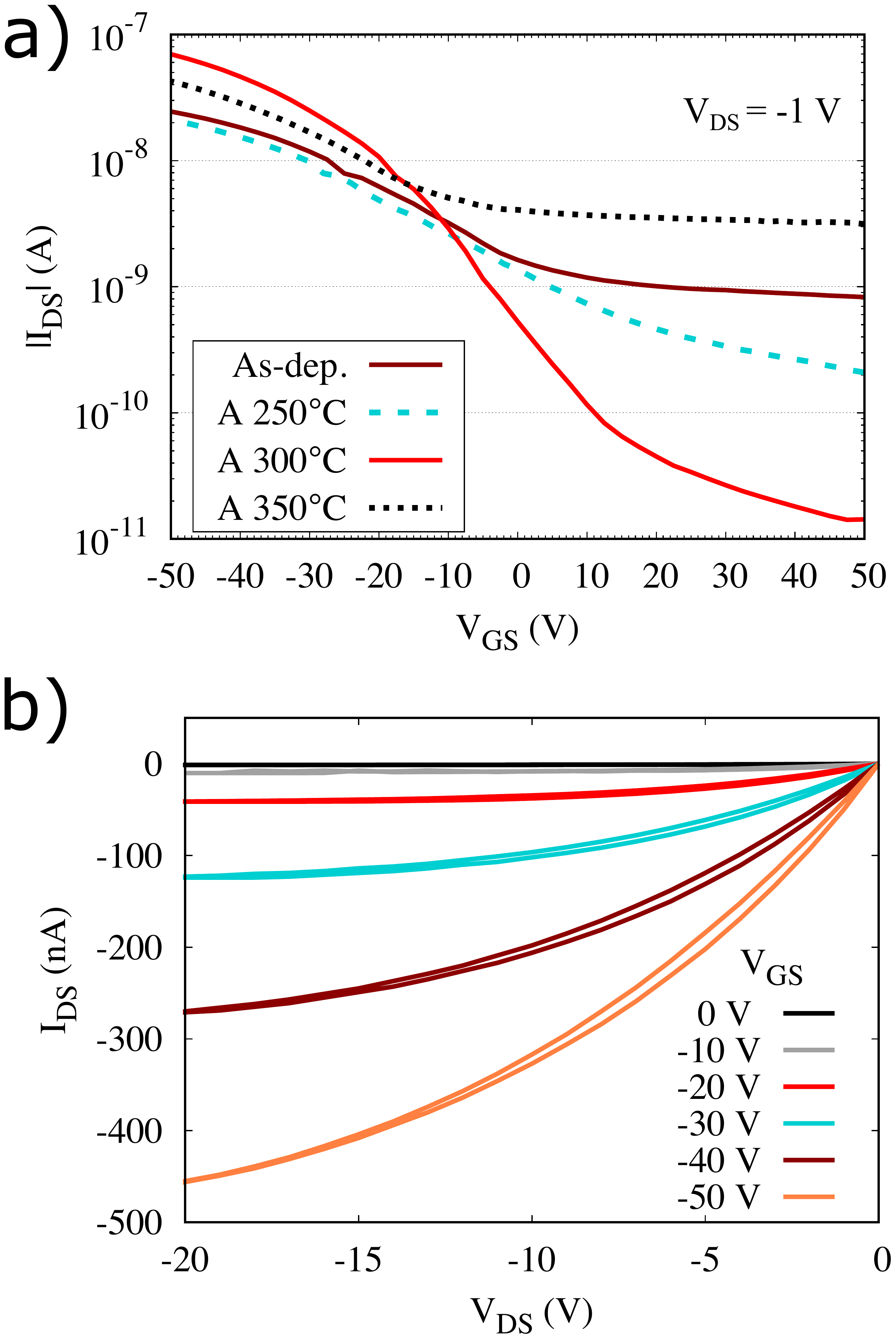}
    \caption{(a) Transfer characteristics of the \ce{Cu2O} TFTs with \ce{Al2O3} passivation, annealed in low-vacuum at different temperatures. Measured with V\textsubscript{DS}=-1.0~V. (b) Output characteristics of a device annealed at 300~$^{\circ}$C.}
    \label{Fig3}
\end{figure}

To further improve TFT performance, the devices were annealed for 10~mins in 1.5~mbar \ce{N2} directly after \ce{Al2O3} deposition. A low vacuum environment was chosen to prevent phase transitions of the \ce{Cu2O} layer into \ce{CuO} or Cu. As seen in Fig \ref{Fig3}(a) the transfer characteristics of the devices started to improve after annealing at 250~$^{\circ}$C, but the most significant effect was gained at 300~$^{\circ}$C, with output characteristics shown in Fig. \ref{Fig3}(b). At higher annealing temperatures transfer characteristics begun to deteriorate. In the devices annealed at 400~$^{\circ}$C no switching was observed, and a positive I\textsubscript{DS} was recorded (data not shown). In the unpatterned \ce{Al2O3}/\ce{Cu2O} film reference sample on glasss the 400~$^{\circ}$C annealing caused color changes visible to the naked eye, potentially indicating a partial reduction into metallic Cu. The same effect was observed also when the annealing was performed in 1~atm Ar atmosphere at the same temperature.

Despite the increase in the switching ratio of up to I\textsubscript{on}/I\textsubscript{off} = $5\cdot 10^3$ in the sample annealed at 300~$^{\circ}$C, the carrier mobility remained low, with field-effect mobility for the as-deposited and annealed devices being \textmu\textsubscript{FE}$\approx 1.5\cdot10^{-3}$~cm$^{2}$V$^{-1}$s$^{-1}$, which was calculated as \textmu\textsubscript{FE}=($g_mL$)/($WC_{\text{ox}}V_{\text{DS}}$), where $g_m$ is the transconductance ($g_m$=$\delta I_{\text{DS}}$/$\delta V_{\text{GS}}$), $L$ and $W$ the channel length and width, respectively, ($L$=50~\textmu m, $W$=1000~\textmu m), and $C$\textsubscript{ox} the gate dielectric capacitance per unit area, calculated using a dielectric constant of 3.9 for \ce{SiO2} gate oxide. Additionally, high operating voltages were required for switching, even for devices with enhanced characteristics, with a threshold voltage $V$\textsubscript{TH} and subthreshold swing SS of -19.8~V and 11.5~V dec$^{-1}$, respectively. The corresponding V\textsubscript{TH} and SS of the as-deposited device with \ce{Al2O3} passivation were -13.0~V and 29.2~V dec$^{-1}$, respectively.

The characteristics of the TFTs without the \ce{Al2O3} layer were not improved upon annealing, and Hall effect measurements showed there to be no change in the carrier density of the annealed \ce{Cu2O} samples (Fig. S1 in the Supporting Information). Therefore, it can be concluded that the \ce{Al2O3} passivation is the reason for the improved performance. Similar effects have been reported for p-type TFTs with passivated SnO channels \cite{Kim2017,Qu2018}. Kim et al. showed an improvement in devices with ALD SnO channel passivated with ALD \ce{Al2O3}, which was further enhanced by subsequent annealing \cite{Kim2017}. Similar observations were made by Qu et al., who passivated sputtered SnO channels by ALD \ce{Al2O3} as well as with organic coatings \cite{Qu2018}. Our results are consistent with these findings, both reporting an increase in the I\textsubscript{on}/I\textsubscript{off} ratio and a decrease in SS upon ALD \ce{Al2O3} passivation. These changes can be associated with a reduction in the trap state density at the channel surface. It seems that the \ce{ALD} \ce{Al2O3} passivation has more impact on reducing the deep trap state density, both in the \ce{Cu2O} and SnO, indicated by the reduction in the SS. On the other hand, the shallow traps (tail states near the valence band), are less affected, as the the carrier mobility does not increase significantly \cite{Qu2018}. Interestingly, it has been reported \cite{Hu2018} that passivation of n-type oxide TFTs by ALD \ce{Al2O3} increases the mobility and SS which is opposite to what has been observed for the p-type devices.

The low field-effect mobility in the order of \textmu\textsubscript{FE} = $10^{-3}-10^{-2}$~cm$^2$V$^{-1}$s$^{-1}$ is typical for \ce{Cu2O} TFTs processed at low/moderate temperatures and with thin channel layer of few tens of nm, regardless of the deposition technique \cite{Wang2016,Fortunato2010,Jang2016}.  However, there are some reports where orders of magnitude higher \textmu\textsubscript{FE} values, up to 6~cm$^2$V$^{-1}$s$^{-1}$ have been achieved, even with a room-temperature processing and mixed phase \ce{Cu2O}-CuO films \cite{Maeng2016,Yao2012}. The limited mobility in \ce{Cu2O} thin films is typically associated with the high density of subgap trap states and grain boundary scattering \cite{Wang2016,Han2017}. Additionally, it has been shown that a CuO layer can form at the \ce{Cu2O}/\ce{SiO2} interface already at 300~$^{\circ}$C, which further increases the trap density and, hence, has a negative impact on the transfer characteristics \cite{Ran2015}. Therefore, it has been suggested that replacing \ce{SiO2} with a high-k dielectric may result in better performance \cite{Zou2010,Zou2011}. 

We also tested devices with a 75~nm thick ALD \ce{Al2O3} gate oxide, and observed switching in the devices with a decreased SS (7.5~V dec$^{-1}$) and a V\textsubscript{TH} of 10.6~V (See Fig. S2), indeed indicating a reduction in the trap states at the dielectric/semiconductor interface. However, in this case the \ce{Al2O3} gate oxide had a lower breakdown voltage than the 100~nm \ce{SiO2}, which meant that the channel was not yet fully depleted when the gate modulation was lost, limiting both the I\textsubscript{on}/I\textsubscript{off} ratio and the mobility.

\begin{figure}
    \centering
    \includegraphics[width=8.5cm]{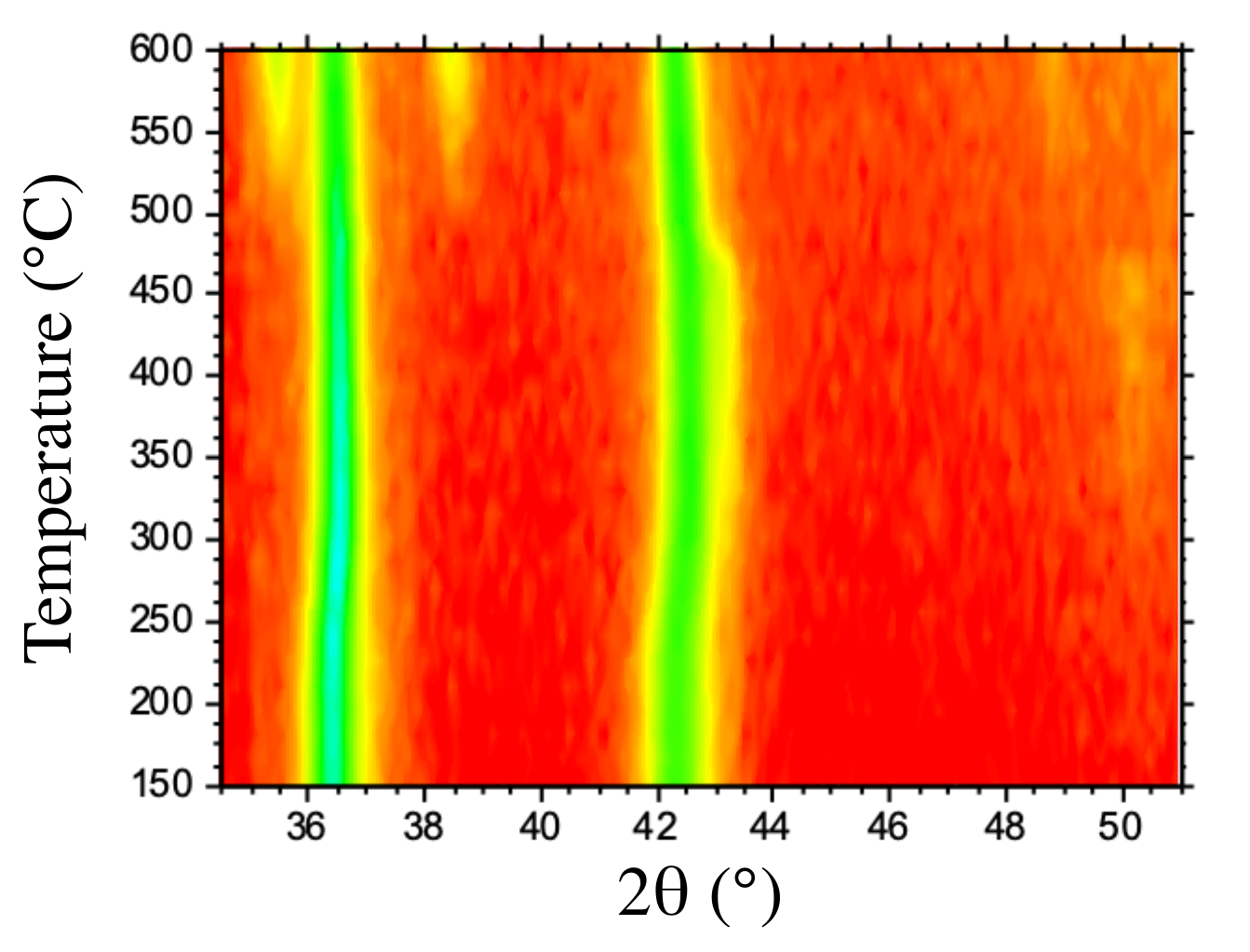}
    \caption{2D high-temperature GIXRD pattern of the 10~nm \ce{Al2O3}/40~nm \ce{Cu2O} stack annealed to 150--600~$^{\circ}$C in 20~mbar \ce{N2}. High intensity lines correspond to the \ce{Cu2O} (111) and (200) reflections. The intensity is plotted in logscale.}
    \label{Fig4}
\end{figure}

To investigate the effect of annealing on the \ce{Al2O3}/\ce{Cu2O} stack in detail, a high-temperature GIXRD measurement was performed. 10~nm \ce{Al2O3} was deposited on a 40~nm \ce{Cu2O} sample and the diffraction patterns were collected at 150--600~$^{\circ}$C in 20~mbar \ce{N2} (Fig. \ref{Fig4}). It was observed that at 250~$^{\circ}$C the (111) and (200) reflections shift towards larger 2\texttheta\ angles, indicating a decrease in the unit-cell parameters, possibly due to stress relaxation when the annealing temperature exceeds the deposition temperature. As seen in the samples annealed earlier, signs of metallic Cu appeared also in the passivated sample. These changes, seen as reflections at ca. 43$^{\circ}$ and 50.5$^{\circ}$, take place just above 300~$^{\circ}$C. At 475$^{\circ}$C the Cu reflections disappear and features corresponding to formation of CuO become visible. These changes in the film structure upon annealing could explain the observed narrow annealing temperature window for optimal TFT performance. 

Moreover, when annealed under similar conditions, a \ce{Cu2O} film without the \ce{Al2O3} layer undergoes oxidation to \ce{CuO} already at 300~$^{\circ}$C (Fig. S3), showing the importance of the \ce{Al2O3} layer to the phase-stability of the films during annealing. This behaviour of both the bare ALD \ce{Cu2O} and the \ce{Al2O3}\ce{Cu2O} film stack differ from what has been shown for \ce{Cu2O} films and devices fabricated by physical deposition methods such as PLD and sputtering. There a high temperature deposition or annealing at 500--800~$^{\circ}$C, both in vacuum and inert gas atmosphere have shown to improve the device performance significantly by reduction of the CuO phases on the grain boundaries and increase in the \ce{Cu2O} crystallite and grain sizes, while the \ce{Cu2O} phase remains stable \cite{Wang2016,Zou2010,Jeong2013,Han2017,Sohn2012,Nam2012}. Though the film thickness may have an effect on the film behaviour during the annealing, it does not fully explain the observed differences between PVD and ALD deposited \ce{Cu2O}. One explanation is that the grain boundaries of the nanocrystalline ALD \ce{Cu2O} contain a higher density of hydroxyl groups, which then accelerate the film reduction, despite the presence of the passivation layer, and the partial oxidation into CuO is later initiated by the oxygen diffusion from both the \ce{Al2O3} layer as well as the \ce{SiO2} gate oxide. However, this remains inconclusive.

\subsection{\ce{Al2O3}/\ce{Cu2O} interface characterisation}

Our results and the previous studies on the passivation of TFTs with oxide semiconductor channels show that quality of the interface between the channel oxide and the passivation layer can have a significant impact on the device performance. Especially, in the case of passivation by chemical routes, such as ALD, it can be assumed that the interface is further modified by the surface chemistry taking place during the layer deposition. It has been shown that exposure to certain ALD metal precursors, namely alkyl compounds, can reduce a surface oxide layer if the reactions are energetically favourable \cite{Lee2012,Lee2014,Gharachorlou2015}. For instance, for surface reactions of diethylzinc, commonly used as a precursor for ALD ZnO, the Gibbs free energies for reduction reactions  of \ce{Cu2O} surface into metallic Cu are \textDelta G\textsubscript{r} = (-300)--(-200)~kJ mol$^{-1}$ (T = 373~K) \cite{Lee2014}. The reactions between TMA and \ce{Cu2O} can be assumed to be even more favourable due to higher reactivity of the TMA. The reduction of oxidised Cu surface during the first \ce{Al2O3} cycles has been verified both numerically and experimentally \cite{Gharachorlou2015,Deuermeier_2016}. Gharachorlou et al. investigated the TMA and hydroxyl-free oxidised copper surface reactions and presented that the TMA is adsorbed and dissociates at the \ce{Cu2O} surface and in those reactions consumes oxygen from the \ce{Cu2O} (or CuO), thus reducing the oxidised Cu\textsuperscript{+}/Cu\textsuperscript{2+} surface to the metallic Cu\textsuperscript{0} state.  Additionally, their results indicated formation of a \ce{CuAlO2} at the interface during the first three \ce{Al2O3} cycles with TMA and \ce{O2}. Their proposed overall reaction is \cite{Gharachorlou2015}:
\begin{equation}
\label{Eq1}
2\ce{Cu2O} + \ce{Al(CH3)3} \rightarrow \ce{CuAlO2} + 3\ce{Cu} + 2\ce{CH4}(g) + \ce{CH}_{ads} \ . 
\end{equation}

Using a process with \ce{H2O} as a reactant leads to surface hydroxyl groups forming during deposition. However, it can be assumed that the mechanism described above is still valid, because it has been calculated that the hydroxyl coverage does not affect the TMA dissociation on the surface, but only on growth efficiency \cite{Elliott2004}.

The formation of a \ce{CuAlO2} interface could be beneficial to TFT performance, because it is a known p-type material with low  V\textsubscript{Cu} formation energy \cite{Kawazoe1997}. In order to investigate in detail the reduction of \ce{Cu2O} by TMA, and the formation of the \ce{CuAlO2} interface layer, samples were prepared for XPS and TEM. XPS was used to analyse \ce{Cu2O} films with and without the \ce{Al2O3} passivation and subsequent annealing at 300~$^{\circ}$C. To minimize the need of Ar$^{+}$ etching to reach the \ce{Al2O3}/\ce{Cu2O} interface, only 20 cycles i.e. ca. 2~nm of \ce{Al2O3} was deposited on samples for XPS measurements. The XPS of as-deposited \ce{Cu2O} without the \ce{Al2O3} layer confirmed the films to be phase-pure \ce{Cu2O} seen both in the Cu 2p and Cu LMM spectra, with a minor CuO content present at the film surface, as well as a high content of hydroxyl groups, as seen in the measured O 1s spectra (See Fig. S4). 

\begin{figure}
    \centering
    \includegraphics[width=8.5cm]{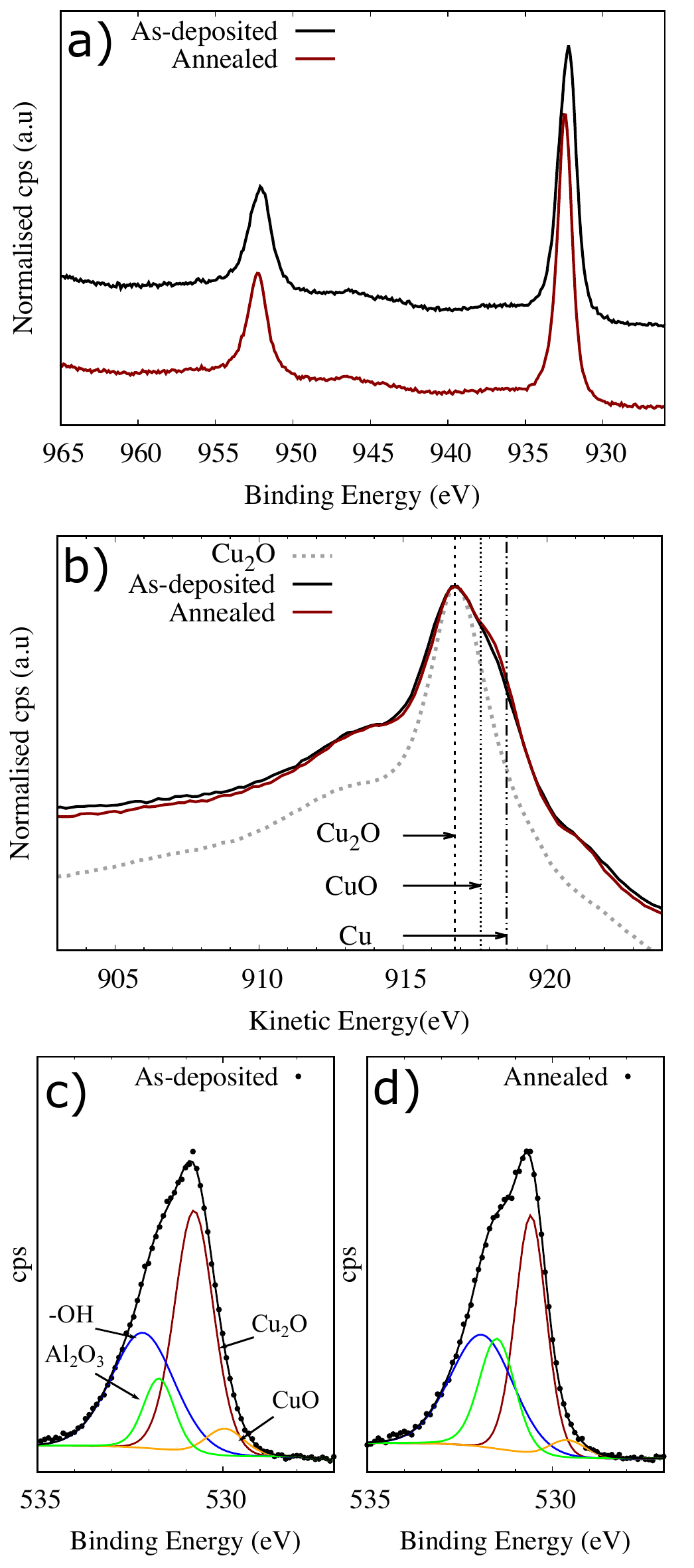}
    \caption{XPS spectra of the \ce{Al2O3}/\ce{Cu2O} interface region, measured after 90~s 0.5~keV Ar$^+$ sputtering, (a) Cu 2p of the as-deposited (black) and annealed (dark red) samples. (b) Cu LMM Auger electron spectra, corresponding spectrum of a \ce{Cu2O} film without the \ce{Al2O3} passivation layer shown by dashed grey line as a reference. The O 1s spectra of the (c) as-deposited and (d) annealed sample. The deconvoluted peaks correspond to oxygen in \ce{Cu2O} (dark red), \ce{CuO} (orange), hydroxyl -OH (blue), and \ce{Al2O3} (green).}
    \label{Fig5}
\end{figure}

The effect of the annealing on the \ce{Cu2O} film was investigated by measuring a sample annealed at 300~$^{\circ}$C. No changes to the film composition were observed (Fig. S6). Figure \ref{Fig5} shows the XPS spectra of the \ce{Al2O3}/\ce{Cu2O} films, before and after annealing at 300~$^{\circ}$C. As seen in Fig. \ref{Fig5}(a) the Cu 2p spectra of the both films show the absence of the CuO phase. However, the differentiation of between the Cu$^{+}$ and Cu$^0$ states cannot be done from the Cu 2p spectrum. The complementary Cu LMM spectra of the samples in Fig. \ref{Fig5}(b) show the significant broadening of the Auger electron peak compared to the \ce{Cu2O} sample without the \ce{Al2O3} layer. This corresponds to  the presence of the metallic Cu$^0$ at the \ce{Al2O3}/\ce{Cu2O} interface, confirming the reduction of the \ce{Cu2O} due to TMA exposure \cite{Biesinger2017}. This Cu$^0$ can remain metallic even after the subsequent pulsing of \ce{H2O}, as the oxidation reactions into \ce{Cu2O} or \ce{CuO} are not thermodynamically favourable (Tab. S1(a) and (b)). However, the oxidation by residual \ce{O2} in the deposition reactor is possible (Tab. S1(b) and (c)).

The O 1s spectra were also recorded from both samples. The deconvoluted spectra in Fig. \ref{Fig5}(c) and (d) show the oxygen in lattice \ce{Cu2O} at 530.2~eV, a minor CuO contribution at 529.8~eV, and \ce{Al2O3} bound oxygen at 531.6~eV, as well as the presence of high hydroxyl concentration ($\sim$532~eV) \cite{Biesinger2010}. This is contributed by both the persistent surface hydroxyl groups on the \ce{Cu2O} as well as the remaining -OH species in the \ce{Al2O3} from the TMA + water process at relatively low deposition temperature of 150~$^{\circ}$C. There is a small difference in the O 1s spectra of the as-deposited and annealed samples, namely in the \ce{Al2O3} related O content, which may indicate a partial diffusion of the Al into the \ce{Cu2O} film or a densification of the film upon annealing, which would lead to a slightly different etching rate during the Ar ion sputter cleaning. 

However, the measured \textit{ex-situ} XPS data can not be reliably used to confirm the presence of the \ce{CuAlO2} phase at the \ce{Al2O3}/\ce{Cu2O} interface, as the related changes are too subtle to be distinguished from the \ce{Al2O3} and \ce{Cu2O} signals within the probed volume. Moreover, we measured the valence spectra of the annealed \ce{Cu2O} and \ce{Al2O3}/\ce{Cu2O} samples. Deuermeier et al. reported a shift in the binding energy of a \ce{Cu2O} during the ALD of \ce{Al2O3}. In their \textit{in situ} XPS experiments on ALD \ce{Al2O3} growth on sputtered \ce{Cu2O} surface the position of the valence band edge ($E_F-E_{VB}$) of the \ce{Cu2O} increased from original 0.4 eV to 0.6 eV after the first \ce{Al2O3} ALD cycle indicating a formation of Cu/\ce{Cu2O} Schottky junction\cite{Deuermeier_2016}. Though our core level LMM spectrum showed the formation of the Cu, similar indication of a Schottky junction formation was not observed in the valence spectra and 0.9~eV $E_F-E_{VB}$ was measured  for both the original \ce{Cu2O} as well as for the \ce{Al2O3}/\ce{Cu2O} interface (See Fig. S5). However, this does not exclude the potential Fermi level pinning at the interface due to the reasons explained above.

\begin{figure*}[ht!]
    \centering
    \includegraphics[width=10cm]{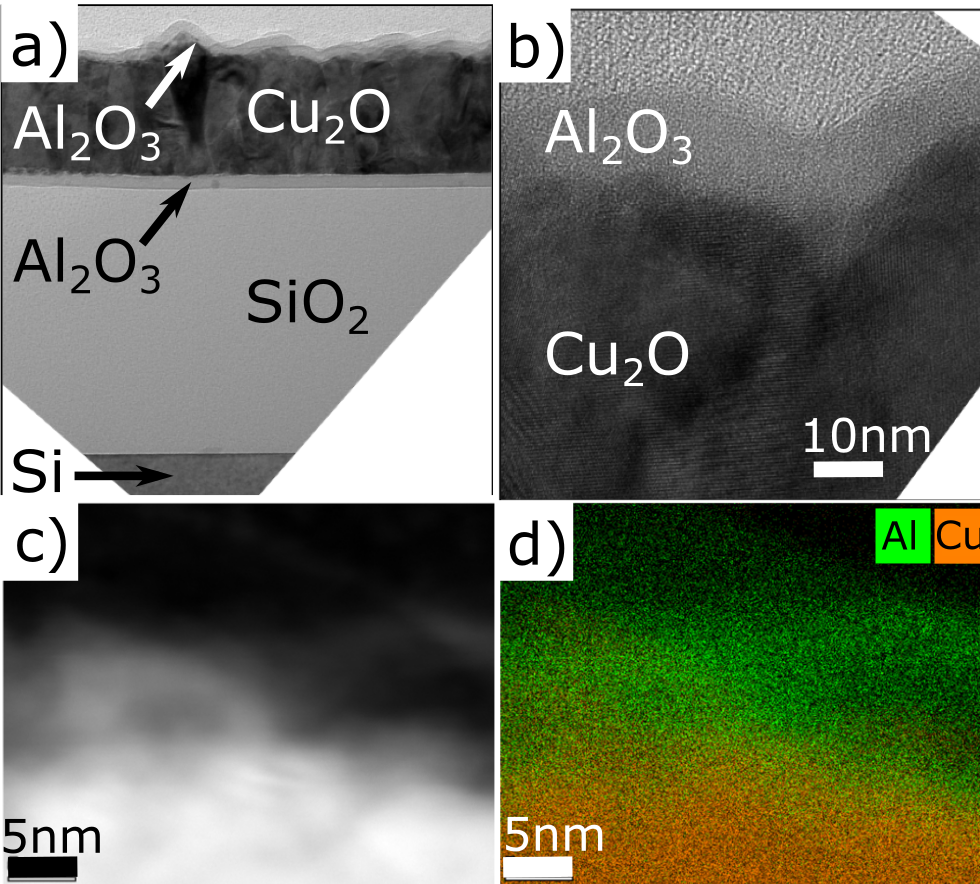}
    \caption{Transmission electron microscopy (TEM) micrographs (a)-(c) of a \ce{Cu2O} thin film sample with 10~nm \ce{Al2O3} passivation layer, annealed in 1.5~mbar \ce{N2} for 10~min. (d) a corresponding TEM-EDS mapping of Al and Cu at the sample interface depicted in (c).}
    \label{Fig6}
\end{figure*}

To obtain more evidence on the proposed \ce{CuAlO2} interface layer formation, a sample with 10~nm \ce{Al2O3} deposited on \ce{Cu2O}, annealed at 300~$^{\circ}$C, was imaged with TEM and TEM-EDS elemental mapping was recorded from the film interface (Fig. \ref{Fig6}).  
Figure \ref{Fig6}(a) shows a low magnification image of the film stack, with a \ce{Cu2O} film sandwiched between two 10~\ce{Al2O3} films on \ce{SiO2}/Si. In Fig. \ref{Fig6}(b) the higher magnification shows the polycrystalline \ce{Cu2O} and with the conformal, amorphous \ce{Al2O3} on top. The EDS measurement of the interface region (\ref{Fig6}(c)) reveals a 1--2~nm region at the interface with a mixed Al and Cu oxide composition (Fig. \ref{Fig6}(d)). Though the actual composition of this region cannot be reliably determined from the TEM-EDS data, it is in qualitative agreement with the observations by Gharachorlou et al. of a formation of a \ce{CuAlO2} layer during the first \ce{Al2O3} cycles according to Eq. \ref{Eq1}.

This reduction mechanism and possible formation of the \ce{CuAlO2} does not fully explain the significant improvement in the TFT performance. We tested the effect of the \ce{Al2O3} film thickness on the TFT transfer characteristics and observed that when the \ce{Al2O3} thickness was only 2--5~nm the device performance was similar to TFTs without the \ce{Al2O3} passivation and the improvement in the TFT transfer characteristics was detected only with a thicker, 10~nm \ce{Al2O3} layers (See Fig. S6). This indicates that the formation of the interface layer by the TMA exposure during the first couple of tens of deposition cycles is not sufficient in improving the device performance, but thicker coverage with the \ce{Al2O3} layer is required. Therefore the key to the improved characteristics are likely in the properties of the ALD \ce{Al2O3} film itself. 

A plausible reason for the observed behaviour is the high negative fixed charge density ($Q$\textsubscript{f} = $10^{-13}$~cm$^{-2}$) of the ALD \ce{Al2O3}, which has traditionally been utilised in c-Si solar cells where it reduces the recombination losses on the Si surface via surface defect density reduction and by field-effect passivation. The field-effect passivation is based on the reduction of electron or hole concentrations on the surface/interface by the means of an intrinsic internal electric field  \cite{Hoex2008,Dingemans2012}. In our measured TFT data the impact of the field-effect passivation is indicated by the negative shift in the V\textsubscript{TH} and the reduction in the I\textsubscript{off}. It has previously been reported by Han et al. that the high off-state current in the \ce{Cu2O} TFTs is due to accumulation of minority charge carriers (electrons) at positive gate voltage regime \cite{Han2017_2}. The application of the field-effect passivation by ALD \ce{Al2O3} on the \ce{Cu2O} channel and subsequent annealing can be an effective way in reducing the accumulation via electrostatic shielding, which leads to the orders of magnitude lower off-state current and, hence, improves the performance of TFTs with \ce{Cu2O} p-channels.

\section{Conclusions}
p-Type thin film transistors with ALD grown phase pure polycrystalline \ce{Cu2O} channel layer were fabricated. The TFTs with as-deposited films showed only limited switching performance, due to the unoptimised film properties and processing parameters, but the characteristics were improved by depositing an 10~nm ALD \ce{Al2O3} passivation layer on the \ce{Cu2O} channel and by subsequent annealing at 300~$^{\circ}$C in low vacuum. The analysis of the transfer characteristics indicates that the improvement is due to the reduced number of trap states at the channel. The detailed investigation of the \ce{Al2O3}/\ce{Cu2O} interface by XPS and TEM showed a partial reduction of the \ce{Cu2O} and possible formation of a 1-2~nm thick \ce{CuAlO2} layer. The p-type \ce{CuAlO2} layer with a low Cu vacancy formation energy can be beneficial to the device operation, but cannot solely explain the better performance of the \ce{Al2O3} passivated TFTs. Hence, we conclude that the main benefit of the \ce{Al2O3} passivation comes from its high negative fixed charge density that reduces the accumulation of electrons in the \ce{Cu2O} channel when positive gate voltage is applied and, thus, reduces the I\textsubscript{off} of the devices. While the field-effect passivation may not be applicable to tradition nanoscale Si-based CMOS devices, as it influences the V\textsubscript{th} and the transport of charge carriers in ways that can be detrimental to the circuit operation, it can be an useful tool in the development of alternative approaches that utilise p-type oxide semiconductors with moderate charge carrier density.

\section{Experimental}

The \ce{Cu2O} films were grown on 5~cm $\times$ 5~cm substrates of thermally grown \ce{SiO2} on p-Si (resistivity 0.001~\textOmega cm, Si-Mat) by atomic layer deposition at 200~$^{\circ}$C in an ASM F-120 reactor. Copper(II) acetate \ce{Cu(OAc)2} with source temperature of 185~$^{\circ}$C, and water vapor were used as precursors. Each \ce{Cu2O} ALD cycle consisted of 2~s \ce{Cu(OAc)2} pulse / 2~s purge / 1.5~s \ce{H2O} pulse / 1.5~s purge, which resulted a growth per cycle of 0.011~nm. A fluorine-free precursor was chosen because residual fluorine impurities can affect the electrical properties of the films, F being a known n-type dopant, and, additionally lead to poor adhesion of the films due to the accumulation of the fluorine into the interfaces \cite{Gandikota2000}. Details of the growth chemistry and materials characterisation are published by Iivonen et al. in Ref. \cite{Iivonen2019}. No further optimisation of the \ce{Cu2O} film processing or thickness was done regarding the device operation. Bottom gate TFT structures, with the Si substrate acting as a common gate and thermally grown 100~nm thick \ce{SiO2} as a gate dielectric, were fabricated with a standard photolithography and nanofabrication methods to test the performance of the \ce{Cu2O} films. The 40~nm \ce{Cu2O} films were patterned by wet etching using diluted (0.025 M aq.) HCl, and 100~nm thick Au source and drain electrodes were deposited by thermal evaporation (Edwards Coating System E306A), with a base pressure of $10^{-6}$~mbar. After electrode patterning, the \ce{Cu2O} channel was passivated by a 10~nm \ce{Al2O3} film grown by ALD (Cambridge Nanotech (Veeco) Savannah S100) at 150~$^{\circ}$C with trimethylaluminium (TMA, Sigma Aldrich) and water vapour. Finally, the devices were annealed in 1.5~mbar \ce{N2} at 200--400~$^{\circ}$C for 10 minutes.

Film thickness was determined with x-ray reflectivity (XRR) using a PANalytical X'Pert Pro MPD diffractometer, which was also used for x-ray diffraction (XRD) measurements. The measurements were performed in the grazing incidence (GIXRD) geometry at an incidence angle of 1$^{\circ}$. The same geometry was used with the high-temperature GIXRD measurements, where Anton-Paar HTK1200N furnace was used for sample heating in 20~mbar \ce{N2} (\ce{N2} flow 40~sccm) and data were collected at 150--600~$^{\circ}C$ with 15~$^{\circ}C$ intervals. Atomic force microscopy (AFM) images were taken with Bruker Multimode 8. The electrical properties of the \ce{Cu2O} films were characterised by Hall-effect measurements using van der Pauw configuration with a magnetic field of 0.2~T at room temperature (MMR Technologies Hall System). Thin film transistors were measured using a Cascade probe station and Agilent B1500A semiconductor device parameter analyser. The electrical characterisations were performed in dark to suppress the film photoconductivity. In the interface examinations Escalab (Thermo Fisher Scientific) x-ray microprobe was used for x-ray photoelectron spectroscopy (XPS), and the results were analysed using CasaXPS processing software. Transmission electron microscopy (TEM) imaging was done using FEI TALOS T200X operated at 200~kV with EDS for elemental mapping.

\begin{acknowledgement}

M.N., T.N.M, A.J.F, and J.L. M.-D. acknowledge funding from the EPSRC grant EP/P027032/1 and PragmatIC Ltd. J.L.M.-D .acknowledges funding from the Royal Academy of Engineering grant CIET 1819\_24. R.L.Z.H. thanks the Royal Academy of Engineering for support via the Research Fellowships scheme (no. RF/201718/1701). Han W. and H.W. acknowledge the support from the U.S. National Science Foundation (NSF, DMR- 2016453) for the TEM work at Purdue University.

\end{acknowledgement}

\begin{suppinfo}

Supporting Information is available from the Wiley Online Library or from the author.

\end{suppinfo}

\bibliography{Cu2O_TFTs.bib}

\providecommand{\latin}[1]{#1}
\makeatletter
\providecommand{\doi}
  {\begingroup\let\do\@makeother\dospecials
  \catcode`\{=1 \catcode`\}=2 \doi@aux}
\providecommand{\doi@aux}[1]{\endgroup\texttt{#1}}
\makeatother
\providecommand*\mcitethebibliography{\thebibliography}
\csname @ifundefined\endcsname{endmcitethebibliography}
  {\let\endmcitethebibliography\endthebibliography}{}
\begin{mcitethebibliography}{49}
\providecommand*\natexlab[1]{#1}
\providecommand*\mciteSetBstSublistMode[1]{}
\providecommand*\mciteSetBstMaxWidthForm[2]{}
\providecommand*\mciteBstWouldAddEndPuncttrue
  {\def\EndOfBibitem{\unskip.}}
\providecommand*\mciteBstWouldAddEndPunctfalse
  {\let\EndOfBibitem\relax}
\providecommand*\mciteSetBstMidEndSepPunct[3]{}
\providecommand*\mciteSetBstSublistLabelBeginEnd[3]{}
\providecommand*\EndOfBibitem{}
\mciteSetBstSublistMode{f}
\mciteSetBstMaxWidthForm{subitem}{(\alph{mcitesubitemcount})}
\mciteSetBstSublistLabelBeginEnd
  {\mcitemaxwidthsubitemform\space}
  {\relax}
  {\relax}

\bibitem[Hosono(2012)]{Hosono2012}
Hosono,~H. In \emph{Handbook of Visual Display Technology}; Chen,~J.,
  Cranton,~W., Fihn,~M., Eds.; Springer Berlin Heidelberg: Berlin, Heidelberg,
  2012; pp 729--749\relax
\mciteBstWouldAddEndPuncttrue
\mciteSetBstMidEndSepPunct{\mcitedefaultmidpunct}
{\mcitedefaultendpunct}{\mcitedefaultseppunct}\relax
\EndOfBibitem
\bibitem[Sheng \latin{et~al.}(2019)Sheng, Hong, Lee, Kim, Sasase, Kim, Hosono,
  and Park]{Sheng2019}
Sheng,~J.; Hong,~T.; Lee,~H.-M.; Kim,~K.; Sasase,~M.; Kim,~J.; Hosono,~H.;
  Park,~J.-S. Amorphous IGZO TFT with High Mobility of $\sim$70 cm2/(V s) via
  Vertical Dimension Control Using PEALD. \emph{ACS Appl. Mater. Inter.}
  \textbf{2019}, \emph{11}, 40300--40309\relax
\mciteBstWouldAddEndPuncttrue
\mciteSetBstMidEndSepPunct{\mcitedefaultmidpunct}
{\mcitedefaultendpunct}{\mcitedefaultseppunct}\relax
\EndOfBibitem
\bibitem[Wang \latin{et~al.}(2016)Wang, Nayak, Caraveo-Frescas, and
  Alshareef]{Wang2016}
Wang,~Z.; Nayak,~P.~K.; Caraveo-Frescas,~J.~A.; Alshareef,~H.~N. Recent
  Developments in p-Type Oxide Semiconductor Materials and Devices. \emph{Adv.
  Mater.} \textbf{2016}, \emph{28}, 3831\relax
\mciteBstWouldAddEndPuncttrue
\mciteSetBstMidEndSepPunct{\mcitedefaultmidpunct}
{\mcitedefaultendpunct}{\mcitedefaultseppunct}\relax
\EndOfBibitem
\bibitem[Nolan and Elliott(2006)Nolan, and Elliott]{Nolan2006}
Nolan,~M.; Elliott,~S.~D. The p-type conduction mechanism in {Cu$_2$O}: a first
  principles study. \emph{Phys. Chem. Chem. Phys.} \textbf{2006}, \emph{8},
  5350--5358\relax
\mciteBstWouldAddEndPuncttrue
\mciteSetBstMidEndSepPunct{\mcitedefaultmidpunct}
{\mcitedefaultendpunct}{\mcitedefaultseppunct}\relax
\EndOfBibitem
\bibitem[Al-Jawhari(2015)]{AlJahwari2015}
Al-Jawhari,~H. A review of recent advances in transparent p-type
  {Cu$_2$O}-based thin film transistors. \emph{Mat. Sci. Semicon. Proc.}
  \textbf{2015}, \emph{40}, 241 -- 252\relax
\mciteBstWouldAddEndPuncttrue
\mciteSetBstMidEndSepPunct{\mcitedefaultmidpunct}
{\mcitedefaultendpunct}{\mcitedefaultseppunct}\relax
\EndOfBibitem
\bibitem[Zhang and Gong(2019)Zhang, and Gong]{Zhang2019}
Zhang,~J.,~N.and~Sun; Gong,~H. Transparent p-Type Semiconductors: Copper-Based
  Oxides and Oxychalcogenides. \emph{Coatings} \textbf{2019}, \emph{9},
  137\relax
\mciteBstWouldAddEndPuncttrue
\mciteSetBstMidEndSepPunct{\mcitedefaultmidpunct}
{\mcitedefaultendpunct}{\mcitedefaultseppunct}\relax
\EndOfBibitem
\bibitem[Matsuzaki \latin{et~al.}(2008)Matsuzaki, Nomura, Yanagi, Kamiya,
  Hirano, and Hosono]{Matsuzaki2008}
Matsuzaki,~K.; Nomura,~K.; Yanagi,~H.; Kamiya,~T.; Hirano,~M.; Hosono,~H.
  Epitaxial growth of high mobility {Cu$_2$O} thin films and application to
  p-channel thin film transistor. \emph{Appl. Phys. Lett.} \textbf{2008},
  \emph{93}, 202107\relax
\mciteBstWouldAddEndPuncttrue
\mciteSetBstMidEndSepPunct{\mcitedefaultmidpunct}
{\mcitedefaultendpunct}{\mcitedefaultseppunct}\relax
\EndOfBibitem
\bibitem[{Zou} \latin{et~al.}(2010){Zou}, {Fang}, {Yuan}, {Li}, {Guan}, and
  {Zhao}]{Zou2010}
{Zou},~X.; {Fang},~G.; {Yuan},~L.; {Li},~M.; {Guan},~W.; {Zhao},~X. Top-Gate
  Low-Threshold Voltage $p\hbox{-}\hbox{Cu}_{2} \hbox{O}$ Thin-Film Transistor
  Grown on $\hbox{SiO}_{2}/ \hbox{Si}$ Substrate Using a High-$\kappa$ HfON
  Gate Dielectric. \emph{IEEE Electron Dev. Lett.} \textbf{2010}, \emph{31},
  827--829\relax
\mciteBstWouldAddEndPuncttrue
\mciteSetBstMidEndSepPunct{\mcitedefaultmidpunct}
{\mcitedefaultendpunct}{\mcitedefaultseppunct}\relax
\EndOfBibitem
\bibitem[Ran \latin{et~al.}(2015)Ran, Taniguti, Hosono, and Kamiya]{Ran2015}
Ran,~F.-Y.; Taniguti,~M.; Hosono,~H.; Kamiya,~T. Analyses of Surface and
  Interfacial Layers in Polycrystalline {Cu$_2$O} Thin-Film Transistors.
  \emph{J. Display Technol.} \textbf{2015}, \emph{11}, 720--724\relax
\mciteBstWouldAddEndPuncttrue
\mciteSetBstMidEndSepPunct{\mcitedefaultmidpunct}
{\mcitedefaultendpunct}{\mcitedefaultseppunct}\relax
\EndOfBibitem
\bibitem[Fortunato \latin{et~al.}(2010)Fortunato, Figueiredo, Barquinha,
  Elamurugu, Barros, Gonçalves, Park, Hwang, and Martins]{Fortunato2010}
Fortunato,~E.; Figueiredo,~V.; Barquinha,~P.; Elamurugu,~E.; Barros,~R.;
  Gonçalves,~G.; Park,~S.-H.~K.; Hwang,~C.-S.; Martins,~R. Thin-film
  transistors based on p-type {Cu$_2$O} thin films produced at room
  temperature. \emph{Appl. Phys. Lett.} \textbf{2010}, \emph{96}, 192102\relax
\mciteBstWouldAddEndPuncttrue
\mciteSetBstMidEndSepPunct{\mcitedefaultmidpunct}
{\mcitedefaultendpunct}{\mcitedefaultseppunct}\relax
\EndOfBibitem
\bibitem[Jeong \latin{et~al.}(2013)Jeong, Sohn, Song, Cho, Lee, Cho, and
  Kwon]{Jeong2013}
Jeong,~C.-Y.; Sohn,~J.; Song,~S.-H.; Cho,~I.-T.; Lee,~J.-H.; Cho,~E.-S.;
  Kwon,~H.-I. Investigation of the charge transport mechanism and subgap
  density of states in p-type {Cu$_2$O} thin-film transistors. \emph{Appl.
  Phys. Lett.} \textbf{2013}, \emph{102}, 082103\relax
\mciteBstWouldAddEndPuncttrue
\mciteSetBstMidEndSepPunct{\mcitedefaultmidpunct}
{\mcitedefaultendpunct}{\mcitedefaultseppunct}\relax
\EndOfBibitem
\bibitem[Han \latin{et~al.}(2016)Han, Niang, Rughoobur, and Flewitt]{Han2016}
Han,~S.; Niang,~K.~M.; Rughoobur,~G.; Flewitt,~A.~J. Effects of post-deposition
  vacuum annealing on film characteristics of p-type {Cu$_2$O} and its impact
  on thin film transistor characteristics. \emph{Appl. Phys. Lett.}
  \textbf{2016}, \emph{109}, 173502\relax
\mciteBstWouldAddEndPuncttrue
\mciteSetBstMidEndSepPunct{\mcitedefaultmidpunct}
{\mcitedefaultendpunct}{\mcitedefaultseppunct}\relax
\EndOfBibitem
\bibitem[Kim \latin{et~al.}(2013)Kim, Ahn, Lee, Kwon, Hwang, Lee, and
  Cho]{Kim2013}
Kim,~S.~Y.; Ahn,~C.~H.; Lee,~J.~H.; Kwon,~Y.~H.; Hwang,~S.; Lee,~J.~Y.;
  Cho,~H.~K. p-Channel Oxide Thin Film Transistors Using Solution-Processed
  Copper Oxide. \emph{ACS Appl. Mater. Inter.} \textbf{2013}, \emph{5},
  2417--2421\relax
\mciteBstWouldAddEndPuncttrue
\mciteSetBstMidEndSepPunct{\mcitedefaultmidpunct}
{\mcitedefaultendpunct}{\mcitedefaultseppunct}\relax
\EndOfBibitem
\bibitem[Jang \latin{et~al.}(2016)Jang, Chung, Kang, and Subramanian]{Jang2016}
Jang,~J.; Chung,~S.; Kang,~H.; Subramanian,~V. P-type {CuO} and {Cu$_2$O}
  transistors derived from a sol–gel copper (II) acetate monohydrate
  precursor. \emph{Thin Solid Films} \textbf{2016}, \emph{600}, 157 --
  161\relax
\mciteBstWouldAddEndPuncttrue
\mciteSetBstMidEndSepPunct{\mcitedefaultmidpunct}
{\mcitedefaultendpunct}{\mcitedefaultseppunct}\relax
\EndOfBibitem
\bibitem[Musselman \latin{et~al.}(2012)Musselman, Marin, Schmidt-Mende, and
  MacManus-Driscoll]{Musselman2012}
Musselman,~K.~P.; Marin,~A.; Schmidt-Mende,~L.; MacManus-Driscoll,~J.~L.
  Incompatible Length Scales in Nanostructured {Cu$_2$O} Solar Cells.
  \emph{Advanced Functional Materials} \textbf{2012}, \emph{22},
  2202--2208\relax
\mciteBstWouldAddEndPuncttrue
\mciteSetBstMidEndSepPunct{\mcitedefaultmidpunct}
{\mcitedefaultendpunct}{\mcitedefaultseppunct}\relax
\EndOfBibitem
\bibitem[Baby \latin{et~al.}(2015)Baby, Garlapati, Dehm, Häming, Kruk, Hahn,
  and Dasgupta]{Baby2015}
Baby,~T.~T.; Garlapati,~S.~K.; Dehm,~S.; Häming,~M.; Kruk,~R.; Hahn,~H.;
  Dasgupta,~S. A General Route toward Complete Room Temperature Processing of
  Printed and High Performance Oxide Electronics. \emph{ACS Nano}
  \textbf{2015}, \emph{9}, 3075--3083\relax
\mciteBstWouldAddEndPuncttrue
\mciteSetBstMidEndSepPunct{\mcitedefaultmidpunct}
{\mcitedefaultendpunct}{\mcitedefaultseppunct}\relax
\EndOfBibitem
\bibitem[{Cho} \latin{et~al.}(2019){Cho}, {Seol}, {Song}, {Choi}, {Song},
  {Yun}, {Chung}, {Bae}, {Park}, and {Jeong}]{Cho2019}
{Cho},~M.~H.; {Seol},~H.; {Song},~A.; {Choi},~S.; {Song},~Y.; {Yun},~P.~S.;
  {Chung},~K.; {Bae},~J.~U.; {Park},~K.; {Jeong},~J.~K. Comparative Study on
  Performance of {IGZO} Transistors With Sputtered and Atomic Layer Deposited
  Channel Layer. \emph{IEEE T. Electron Dev.} \textbf{2019}, \emph{66},
  1783--1788\relax
\mciteBstWouldAddEndPuncttrue
\mciteSetBstMidEndSepPunct{\mcitedefaultmidpunct}
{\mcitedefaultendpunct}{\mcitedefaultseppunct}\relax
\EndOfBibitem
\bibitem[Sheng \latin{et~al.}(2018)Sheng, Lee, Choi, Hong, Kim, and
  Park]{Sheng2018}
Sheng,~J.; Lee,~J.-H.; Choi,~W.-H.; Hong,~T.; Kim,~M.; Park,~J.-S. Review
  Article: Atomic layer deposition for oxide semiconductor thin film
  transistors: Advances in research and development. \emph{J. Vac. Sci.
  Technol. A} \textbf{2018}, \emph{36}, 060801\relax
\mciteBstWouldAddEndPuncttrue
\mciteSetBstMidEndSepPunct{\mcitedefaultmidpunct}
{\mcitedefaultendpunct}{\mcitedefaultseppunct}\relax
\EndOfBibitem
\bibitem[Sheng \latin{et~al.}(2018)Sheng, Han, Hong, Choi, and
  Park]{Sheng2018_2}
Sheng,~J.; Han,~K.-L.; Hong,~T.; Choi,~W.-H.; Park,~J.-S. Review of recent
  progresses on flexible oxide semiconductor thin film transistors based on
  atomic layer deposition processes. \emph{J. Semicond.} \textbf{2018},
  \emph{39}, 011008\relax
\mciteBstWouldAddEndPuncttrue
\mciteSetBstMidEndSepPunct{\mcitedefaultmidpunct}
{\mcitedefaultendpunct}{\mcitedefaultseppunct}\relax
\EndOfBibitem
\bibitem[Zardetto \latin{et~al.}(2017)Zardetto, Williams, Perrotta, Di~Giacomo,
  Verheijen, Andriessen, Kessels, and Creatore]{Zardetto2017}
Zardetto,~V.; Williams,~B.~L.; Perrotta,~A.; Di~Giacomo,~F.; Verheijen,~M.~A.;
  Andriessen,~R.; Kessels,~W. M.~M.; Creatore,~M. Atomic layer deposition for
  perovskite solar cells: research status{,} opportunities and challenges.
  \emph{Sustain. Energy Fuels} \textbf{2017}, \emph{1}, 30--55\relax
\mciteBstWouldAddEndPuncttrue
\mciteSetBstMidEndSepPunct{\mcitedefaultmidpunct}
{\mcitedefaultendpunct}{\mcitedefaultseppunct}\relax
\EndOfBibitem
\bibitem[Maeng \latin{et~al.}(2016)Maeng, Lee, Kwon, Park, and Park]{Maeng2016}
Maeng,~W.; Lee,~S.-H.; Kwon,~J.-D.; Park,~J.; Park,~J.-S. Atomic layer
  deposited p-type copper oxide thin films and the associated thin film
  transistor properties. \emph{Ceram. Int.} \textbf{2016}, \emph{42}, 5517 --
  5522\relax
\mciteBstWouldAddEndPuncttrue
\mciteSetBstMidEndSepPunct{\mcitedefaultmidpunct}
{\mcitedefaultendpunct}{\mcitedefaultseppunct}\relax
\EndOfBibitem
\bibitem[Kim \latin{et~al.}(2017)Kim, Baek, Kim, Pyeon, Chung, Baek, Kim, Han,
  and Kim]{Kim2017}
Kim,~S.~H.; Baek,~I.-H.; Kim,~D.~H.; Pyeon,~J.~J.; Chung,~T.-M.; Baek,~S.-H.;
  Kim,~J.-S.; Han,~J.~H.; Kim,~S.~K. Fabrication of high-performance p-type
  thin film transistors using atomic-layer-deposited {SnO} films. \emph{J.
  Mater. Chem. C} \textbf{2017}, \emph{5}, 3139--3145\relax
\mciteBstWouldAddEndPuncttrue
\mciteSetBstMidEndSepPunct{\mcitedefaultmidpunct}
{\mcitedefaultendpunct}{\mcitedefaultseppunct}\relax
\EndOfBibitem
\bibitem[Iivonen \latin{et~al.}(2019)Iivonen, Heikkilä, Popov, Nieminen,
  Kaipio, Kemell, Mattinen, Meinander, Mizohata, Räisänen, Ritala, and
  Leskelä]{Iivonen2019}
Iivonen,~T.; Heikkilä,~M.~J.; Popov,~G.; Nieminen,~H.-E.; Kaipio,~M.;
  Kemell,~M.; Mattinen,~M.; Meinander,~K.; Mizohata,~K.; Räisänen,~J.;
  Ritala,~M.; Leskelä,~M. Atomic Layer Deposition of Photoconductive {Cu$_2$O}
  Thin Films. \emph{ACS Omega} \textbf{2019}, \emph{4}, 11205--11214\relax
\mciteBstWouldAddEndPuncttrue
\mciteSetBstMidEndSepPunct{\mcitedefaultmidpunct}
{\mcitedefaultendpunct}{\mcitedefaultseppunct}\relax
\EndOfBibitem
\bibitem[Muñoz-Rojas \latin{et~al.}(2012)Muñoz-Rojas, Jordan, Yeoh, Marin,
  Kursumovic, Dunlop, Iza, Chen, Wang, and MacManus~Driscoll]{Munos-Rojas2012}
Muñoz-Rojas,~D.; Jordan,~M.; Yeoh,~C.; Marin,~A.~T.; Kursumovic,~A.;
  Dunlop,~L.~A.; Iza,~D.~C.; Chen,~A.; Wang,~H.; MacManus~Driscoll,~J.~L.
  Growth of $\sim$5 cm2V-1s-1 mobility, p-type Copper(I) oxide ({Cu$_2$O})
  films by fast atmospheric atomic layer deposition ({AALD}) at 225C and below.
  \emph{AIP Adv.} \textbf{2012}, \emph{2}, 042179\relax
\mciteBstWouldAddEndPuncttrue
\mciteSetBstMidEndSepPunct{\mcitedefaultmidpunct}
{\mcitedefaultendpunct}{\mcitedefaultseppunct}\relax
\EndOfBibitem
\bibitem[Chen \latin{et~al.}(2014)Chen, Hsu, Chien, Chang, Hsu, Chang, Lee,
  Chou, Hsieh, and Wu]{Chen2014}
Chen,~W.-C.; Hsu,~P.-C.; Chien,~C.-W.; Chang,~K.-M.; Hsu,~C.-J.; Chang,~C.-H.;
  Lee,~W.-K.; Chou,~W.-F.; Hsieh,~H.-H.; Wu,~C.-C. Room-temperature-processed
  flexible n-{InGaZnO}/p-{Cu$_2$O} heterojunction diodes and high-frequency
  diode rectifiers. \emph{J. Phys. D: Appl. Phys.} \textbf{2014}, \emph{47},
  365101\relax
\mciteBstWouldAddEndPuncttrue
\mciteSetBstMidEndSepPunct{\mcitedefaultmidpunct}
{\mcitedefaultendpunct}{\mcitedefaultseppunct}\relax
\EndOfBibitem
\bibitem[Han and Flewitt(2017)Han, and Flewitt]{Han2017}
Han,~S.; Flewitt,~A.~J. Analysis of the Conduction Mechanism and Copper Vacancy
  Density in p-type {Cu$_2$O} Thin Films. \emph{Sci. Rep.} \textbf{2017},
  \emph{7}, 5766\relax
\mciteBstWouldAddEndPuncttrue
\mciteSetBstMidEndSepPunct{\mcitedefaultmidpunct}
{\mcitedefaultendpunct}{\mcitedefaultseppunct}\relax
\EndOfBibitem
\bibitem[Deuermeier \latin{et~al.}(2018)Deuermeier, Liu, Rapenne, Calmeiro,
  Renou, Martins, Muñoz-Rojas, and Fortunato]{Deuermeier2018}
Deuermeier,~J.; Liu,~H.; Rapenne,~L.; Calmeiro,~T.; Renou,~G.; Martins,~R.;
  Muñoz-Rojas,~D.; Fortunato,~E. Visualization of nanocrystalline {CuO} in the
  grain boundaries of {Cu$_2$O} thin films and effect on band bending and film
  resistivity. \emph{APL Materials} \textbf{2018}, \emph{6}, 096103\relax
\mciteBstWouldAddEndPuncttrue
\mciteSetBstMidEndSepPunct{\mcitedefaultmidpunct}
{\mcitedefaultendpunct}{\mcitedefaultseppunct}\relax
\EndOfBibitem
\bibitem[Chen \latin{et~al.}(2010)Chen, Chang, Li, Chen, Lu, Chung, Tai, and
  Tseng]{Chen2010}
Chen,~Y.-C.; Chang,~T.-C.; Li,~H.-W.; Chen,~S.-C.; Lu,~J.; Chung,~W.-F.;
  Tai,~Y.-H.; Tseng,~T.-Y. Bias-induced oxygen adsorption in zinc tin oxide
  thin film transistors under dynamic stress. \emph{Appl. Phys. Lett.}
  \textbf{2010}, \emph{96}, 262104\relax
\mciteBstWouldAddEndPuncttrue
\mciteSetBstMidEndSepPunct{\mcitedefaultmidpunct}
{\mcitedefaultendpunct}{\mcitedefaultseppunct}\relax
\EndOfBibitem
\bibitem[{Hu} \latin{et~al.}(2018){Hu}, {Ning}, {Lu}, {Fang}, {Tao}, {Yao},
  {Zou}, {Xu}, {Wang}, and {Peng}]{Hu2018}
{Hu},~S.; {Ning},~H.; {Lu},~K.; {Fang},~Z.; {Tao},~R.; {Yao},~R.; {Zou},~J.;
  {Xu},~M.; {Wang},~L.; {Peng},~J. Effect of {Al$_2$O$_3$} Passivation Layer
  and Cu Electrodes on High Mobility of Amorphous {IZO} {TFT}. \emph{IEEE J.
  Electron Devi.} \textbf{2018}, \emph{6}, 733--737\relax
\mciteBstWouldAddEndPuncttrue
\mciteSetBstMidEndSepPunct{\mcitedefaultmidpunct}
{\mcitedefaultendpunct}{\mcitedefaultseppunct}\relax
\EndOfBibitem
\bibitem[Hong \latin{et~al.}(2017)Hong, Park, Kim, Kang, Na, and Kim]{Hong2017}
Hong,~S.; Park,~S.~P.; Kim,~Y.-g.; Kang,~B.~H.; Na,~J.~W.; Kim,~H.~J.
  Low-temperature fabrication of an {HfO$_2$} passivation layer for amorphous
  indium-gallium-zinc oxide thin film transistors using a solution process.
  \emph{Sci. Rep.} \textbf{2017}, \emph{7}, 16265\relax
\mciteBstWouldAddEndPuncttrue
\mciteSetBstMidEndSepPunct{\mcitedefaultmidpunct}
{\mcitedefaultendpunct}{\mcitedefaultseppunct}\relax
\EndOfBibitem
\bibitem[Tak \latin{et~al.}(2020)Tak, Keene, Kang, Kim, Kim, Salleo, and
  Kim]{Tak2020}
Tak,~Y.~J.; Keene,~S.~T.; Kang,~B.~H.; Kim,~W.-G.; Kim,~S.~J.; Salleo,~A.;
  Kim,~H.~J. Multifunctional, Room-Temperature Processable, Heterogeneous
  Organic Passivation Layer for Oxide Semiconductor Thin-Film Transistors.
  \emph{ACS Appl. Mater. Inter.} \textbf{2020}, \emph{12}, 2615--2624\relax
\mciteBstWouldAddEndPuncttrue
\mciteSetBstMidEndSepPunct{\mcitedefaultmidpunct}
{\mcitedefaultendpunct}{\mcitedefaultseppunct}\relax
\EndOfBibitem
\bibitem[Qu \latin{et~al.}(2018)Qu, Yang, Li, Zhang, Wang, Song, and
  Xin]{Qu2018}
Qu,~Y.; Yang,~J.; Li,~Y.; Zhang,~J.; Wang,~Q.; Song,~A.; Xin,~Q. Organic and
  inorganic passivation of p-type {SnO} thin-film transistors with different
  active layer thicknesses. \emph{Semicond. Sci. Tech.} \textbf{2018},
  \emph{33}, 075001\relax
\mciteBstWouldAddEndPuncttrue
\mciteSetBstMidEndSepPunct{\mcitedefaultmidpunct}
{\mcitedefaultendpunct}{\mcitedefaultseppunct}\relax
\EndOfBibitem
\bibitem[Yao \latin{et~al.}(2012)Yao, Liu, Zhang, He, Kumar, Jiang, Zhang, and
  Shao]{Yao2012}
Yao,~Z.~Q.; Liu,~S.~L.; Zhang,~L.; He,~B.; Kumar,~A.; Jiang,~X.; Zhang,~W.~J.;
  Shao,~G. Room temperature fabrication of p-channel {Cu$_2$O} thin-film
  transistors on flexible polyethylene terephthalate substrates. \emph{Appl.
  Phys. Lett.} \textbf{2012}, \emph{101}, 042114\relax
\mciteBstWouldAddEndPuncttrue
\mciteSetBstMidEndSepPunct{\mcitedefaultmidpunct}
{\mcitedefaultendpunct}{\mcitedefaultseppunct}\relax
\EndOfBibitem
\bibitem[{Zou} \latin{et~al.}(2011){Zou}, {Fang}, {Wan}, {He}, {Wang}, {Liu},
  {Long}, and {Zhao}]{Zou2011}
{Zou},~X.; {Fang},~G.; {Wan},~J.; {He},~X.; {Wang},~H.; {Liu},~N.; {Long},~H.;
  {Zhao},~X. Improved Subthreshold Swing and Gate-Bias Stressing Stability of
  p-Type $\hbox{Cu}_{2}\hbox{O}$ Thin-Film Transistors Using a $\hbox{HfO}_{2}$
  High- $k$ Gate Dielectric Grown on a $\hbox{SiO}_{2}/\hbox{Si}$ Substrate by
  Pulsed Laser Ablation. \emph{IEEE T. Electron Dev.} \textbf{2011}, \emph{58},
  2003--2007\relax
\mciteBstWouldAddEndPuncttrue
\mciteSetBstMidEndSepPunct{\mcitedefaultmidpunct}
{\mcitedefaultendpunct}{\mcitedefaultseppunct}\relax
\EndOfBibitem
\bibitem[Sohn \latin{et~al.}(2012)Sohn, Song, Nam, Cho, Cho, Lee, and
  Kwon]{Sohn2012}
Sohn,~J.; Song,~S.-H.; Nam,~D.-W.; Cho,~I.-T.; Cho,~E.-S.; Lee,~J.-H.;
  Kwon,~H.-I. Effects of vacuum annealing on the optical and electrical
  properties of p-type copper-oxide thin-film transistors. \emph{Semicond. Sci.
  Tech.} \textbf{2012}, \emph{28}, 015005\relax
\mciteBstWouldAddEndPuncttrue
\mciteSetBstMidEndSepPunct{\mcitedefaultmidpunct}
{\mcitedefaultendpunct}{\mcitedefaultseppunct}\relax
\EndOfBibitem
\bibitem[Nam \latin{et~al.}(2012)Nam, Cho, Lee, Cho, Sohn, Song, and
  Kwon]{Nam2012}
Nam,~D.-W.; Cho,~I.-T.; Lee,~J.-H.; Cho,~E.-S.; Sohn,~J.; Song,~S.-H.;
  Kwon,~H.-I. Active layer thickness effects on the structural and electrical
  properties of p-type {Cu$_2$O} thin-film transistors. \emph{J. Vac. Sci.
  Technol. B} \textbf{2012}, \emph{30}, 060605\relax
\mciteBstWouldAddEndPuncttrue
\mciteSetBstMidEndSepPunct{\mcitedefaultmidpunct}
{\mcitedefaultendpunct}{\mcitedefaultseppunct}\relax
\EndOfBibitem
\bibitem[Lee \latin{et~al.}(2012)Lee, Liu, Heo, and Gordon]{Lee2012}
Lee,~S.~W.; Liu,~Y.; Heo,~J.; Gordon,~R.~G. Creation and Control of
  Two-Dimensional Electron Gas Using Al-Based Amorphous Oxides/SrTiO3
  Heterostructures Grown by Atomic Layer Deposition. \emph{Nano Letters}
  \textbf{2012}, \emph{12}, 4775--4783\relax
\mciteBstWouldAddEndPuncttrue
\mciteSetBstMidEndSepPunct{\mcitedefaultmidpunct}
{\mcitedefaultendpunct}{\mcitedefaultseppunct}\relax
\EndOfBibitem
\bibitem[Lee \latin{et~al.}(2014)Lee, Lee, Heo, Siah, Chua, Brandt, Kim,
  Mailoa, Buonassisi, and Gordon]{Lee2014}
Lee,~S.~W.; Lee,~Y.~S.; Heo,~J.; Siah,~S.~C.; Chua,~D.; Brandt,~R.~E.;
  Kim,~S.~B.; Mailoa,~J.~P.; Buonassisi,~T.; Gordon,~R.~G. Improved
  {Cu$_2$O}-Based Solar Cells Using Atomic Layer Deposition to Control the Cu
  Oxidation State at the p-n Junction. \emph{Adv. Energy Mater.} \textbf{2014},
  \emph{4}, 1301916\relax
\mciteBstWouldAddEndPuncttrue
\mciteSetBstMidEndSepPunct{\mcitedefaultmidpunct}
{\mcitedefaultendpunct}{\mcitedefaultseppunct}\relax
\EndOfBibitem
\bibitem[Gharachorlou \latin{et~al.}(2015)Gharachorlou, Detwiler, Gu, Mayr,
  Klötzer, Greeley, Reifenberger, Delgass, Ribeiro, and
  Zemlyanov]{Gharachorlou2015}
Gharachorlou,~A.; Detwiler,~M.~D.; Gu,~X.-K.; Mayr,~L.; Klötzer,~B.;
  Greeley,~J.; Reifenberger,~R.~G.; Delgass,~W.~N.; Ribeiro,~F.~H.;
  Zemlyanov,~D.~Y. Trimethylaluminum and Oxygen Atomic Layer Deposition on
  Hydroxyl-Free {Cu(111)}. \emph{ACS Appl. Mater. Inter.} \textbf{2015},
  \emph{7}, 16428--16439\relax
\mciteBstWouldAddEndPuncttrue
\mciteSetBstMidEndSepPunct{\mcitedefaultmidpunct}
{\mcitedefaultendpunct}{\mcitedefaultseppunct}\relax
\EndOfBibitem
\bibitem[Deuermeier \latin{et~al.}(2016)Deuermeier, Bayer, Yanagi, Kiazadeh,
  Martins, Klein, and Fortunato]{Deuermeier_2016}
Deuermeier,~J.; Bayer,~T. J.~M.; Yanagi,~H.; Kiazadeh,~A.; Martins,~R.;
  Klein,~A.; Fortunato,~E. Substrate reactivity as the origin of Fermi level
  pinning at the {Cu$_2$O}/{ALD}-{Al$_2$O$_3$} interface. \emph{Mater. Res.
  Express} \textbf{2016}, \emph{3}, 046404\relax
\mciteBstWouldAddEndPuncttrue
\mciteSetBstMidEndSepPunct{\mcitedefaultmidpunct}
{\mcitedefaultendpunct}{\mcitedefaultseppunct}\relax
\EndOfBibitem
\bibitem[Elliott and Greer(2004)Elliott, and Greer]{Elliott2004}
Elliott,~S.~D.; Greer,~J.~C. Simulating the atomic layer deposition of alumina
  from first principles. \emph{J. Mater. Chem.} \textbf{2004}, \emph{14},
  3246--3250\relax
\mciteBstWouldAddEndPuncttrue
\mciteSetBstMidEndSepPunct{\mcitedefaultmidpunct}
{\mcitedefaultendpunct}{\mcitedefaultseppunct}\relax
\EndOfBibitem
\bibitem[Kawazoe \latin{et~al.}(1997)Kawazoe, Yasukawa, Hyodo, Kurita, Yanagi,
  and Hosono]{Kawazoe1997}
Kawazoe,~H.; Yasukawa,~M.; Hyodo,~H.; Kurita,~M.; Yanagi,~H.; Hosono,~H. P-type
  electrical conduction in transparent thin films of {CuAlO$_2$}. \emph{Nature}
  \textbf{1997}, \emph{389}, 939--942\relax
\mciteBstWouldAddEndPuncttrue
\mciteSetBstMidEndSepPunct{\mcitedefaultmidpunct}
{\mcitedefaultendpunct}{\mcitedefaultseppunct}\relax
\EndOfBibitem
\bibitem[Biesinger(2017)]{Biesinger2017}
Biesinger,~M.~C. Advanced analysis of copper X-ray photoelectron spectra.
  \emph{Surf. Interface Anal.} \textbf{2017}, \emph{49}, 1325--1334\relax
\mciteBstWouldAddEndPuncttrue
\mciteSetBstMidEndSepPunct{\mcitedefaultmidpunct}
{\mcitedefaultendpunct}{\mcitedefaultseppunct}\relax
\EndOfBibitem
\bibitem[Biesinger \latin{et~al.}(2010)Biesinger, Lau, Gerson, and
  Smart]{Biesinger2010}
Biesinger,~M.~C.; Lau,~L.~W.; Gerson,~A.~R.; Smart,~R.~S. Resolving surface
  chemical states in {XPS} analysis of first row transition metals, oxides and
  hydroxides: Sc, Ti, V, Cu and Zn. \emph{Appl. Surf. Sci.} \textbf{2010},
  \emph{257}, 887 -- 898\relax
\mciteBstWouldAddEndPuncttrue
\mciteSetBstMidEndSepPunct{\mcitedefaultmidpunct}
{\mcitedefaultendpunct}{\mcitedefaultseppunct}\relax
\EndOfBibitem
\bibitem[Hoex \latin{et~al.}(2008)Hoex, Gielis, van~de Sanden, and
  Kessels]{Hoex2008}
Hoex,~B.; Gielis,~J. J.~H.; van~de Sanden,~M. C.~M.; Kessels,~W. M.~M. On the
  {c-Si} surface passivation mechanism by the negative-charge-dielectric
  {Al$_2$O$_3$}. \emph{J. Appl. Phys.} \textbf{2008}, \emph{104}, 113703\relax
\mciteBstWouldAddEndPuncttrue
\mciteSetBstMidEndSepPunct{\mcitedefaultmidpunct}
{\mcitedefaultendpunct}{\mcitedefaultseppunct}\relax
\EndOfBibitem
\bibitem[Dingemans and Kessels(2012)Dingemans, and Kessels]{Dingemans2012}
Dingemans,~G.; Kessels,~W. M.~M. Status and prospects of {Al$_2$O$_3$}-based
  surface passivation schemes for silicon solar cells. \emph{J. Vac. Sci.
  Technol. A} \textbf{2012}, \emph{30}, 040802\relax
\mciteBstWouldAddEndPuncttrue
\mciteSetBstMidEndSepPunct{\mcitedefaultmidpunct}
{\mcitedefaultendpunct}{\mcitedefaultseppunct}\relax
\EndOfBibitem
\bibitem[{Han} and {Flewitt}(2017){Han}, and {Flewitt}]{Han2017_2}
{Han},~S.; {Flewitt},~A.~J. The Origin of the High Off-State Current in p-Type
  {Cu$_2$O} Thin Film Transistors. \emph{IEEE Electron Dev. Lett.}
  \textbf{2017}, \emph{38}, 1394--1397\relax
\mciteBstWouldAddEndPuncttrue
\mciteSetBstMidEndSepPunct{\mcitedefaultmidpunct}
{\mcitedefaultendpunct}{\mcitedefaultseppunct}\relax
\EndOfBibitem
\bibitem[Gandikota \latin{et~al.}(2000)Gandikota, Voss, Tao, Duboust, Cong,
  Chen, Ramaswami, and Carl]{Gandikota2000}
Gandikota,~S.; Voss,~S.; Tao,~R.; Duboust,~A.; Cong,~D.; Chen,~L.-Y.;
  Ramaswami,~S.; Carl,~D. Adhesion studies of CVD copper metallization.
  \emph{Microelectron. Eng.} \textbf{2000}, \emph{50}, 547 -- 553\relax
\mciteBstWouldAddEndPuncttrue
\mciteSetBstMidEndSepPunct{\mcitedefaultmidpunct}
{\mcitedefaultendpunct}{\mcitedefaultseppunct}\relax
\EndOfBibitem
\end{mcitethebibliography}

\end{document}